\documentclass{ws-ijmpa}
\usepackage[utf8x]{inputenc}
\usepackage[T1]{fontenc}
\usepackage[super,compress]{cite}
\usepackage{graphicx}
\usepackage[colorlinks=true]{hyperref}
\usepackage{fdsymbol}

\newcommand{\vt}{\boldsymbol{\mathfrak{t}}}
\newcommand{\vs}{\boldsymbol{\mathfrak{s}}}
\newcommand{\pdt}{\boldsymbol{\partial}_{\vt}}
\newcommand{\dd}{\mathrm{d}}
\newcommand{\DD}{\mathrm{D}}
\newcommand{\ddd}{\boldsymbol{\dd}}
\newcommand{\DDD}{\boldsymbol{\DD}}
\newcommand{\Lie}{\pounds}

\newcommand{\TT}{\boldsymbol{\mathcal{T}}}
\newcommand{\AT}{\boldsymbol{\mathcal{A}}}
\newcommand{\VT}{\boldsymbol{\mathcal{V}}}
\newcommand{\vek}[1]{\overset{\odot}{#1}\vphantom{#1}}
\newcommand{\asy}[1]{\overset{\ominus}{#1}\vphantom{#1}}
\newcommand{\sym}[1]{\overset{\oplus}{#1}\vphantom{#1}}
\newcommand{\trc}[1]{\overset{\otimes}{#1}\vphantom{#1}}

\begin{document}
\markboth{Manuel Hohmann}{Hamiltonian of new general relativity using differential forms}

%
\catchline{}{}{}{}{}
%

\title{Hamiltonian of new general relativity using differential forms}

\author{Manuel Hohmann}

\address{Laboratory of Theoretical Physics, Institute of Physics, University of Tartu\\
W. Ostwaldi 1, 50411 Tartu, Estonia\\
manuel.hohmann@ut.ee}

\maketitle

\begin{history}
\received{Day Month Year}
\revised{Day Month Year}
\end{history}

\begin{abstract}
In a recent work we derived the kinematic Hamiltonian and primary constraints of the new general relativity class of teleparallel gravity theories and showed that these theories can be grouped in 9 classes, based on the presence or absence of primary constraints in their Hamiltonian. Here we demonstrate an alternative approach towards this result, by using differential forms instead of tensor components throughout the calculation. We prove that also this alternative derivation yields the same results and show how they are related to each other.

\end{abstract}

\ccode{PACS numbers: 04.50.Kd, 47.10.Df}


\section{Introduction}
We recently derived the kinematic Hamiltonian and primary constraints of new general relativity\cite{Hayashi:1979qx} using the language of tensor components.\cite{Blixt:2018znp,Blixt:2019ene} For two particular examples from this class of theories, a toy model\cite{Okolow:2011np} as well as the teleparallel equivalent of general relativity,\cite{Okolow:2011nq,Okolow:2013lwa} the Hamiltonian has also been derived using differential forms. Here we demonstrate how the kinematic Hamiltonian and primary constraints are derived in this latter formalism in the general case.

\section{Geometric setting}
We assume a globally hyperbolic spacetime manifold \(M \cong \mathbb{R} \times \Sigma\), on which the dynamical fields are given by the coframe \(\boldsymbol{\theta}^a \in \Omega^1(M)\) with \(a = 0, \ldots, 3\) and spin connection \(\boldsymbol{\omega}^a{}_b \in \Omega^1(M)\). They define the metric \(\mathbf{g}\) and the torsion \(\mathbf{T}^a\) via
\begin{equation}
\mathbf{g} = \eta_{ab}\boldsymbol{\theta}^a \otimes \boldsymbol{\theta}^b\,, \quad
\mathbf{T}^a = \DDD\boldsymbol{\theta}^a = \ddd\boldsymbol{\theta}^a + \boldsymbol{\omega}^a{}_b \wedge \boldsymbol{\theta}^b\,.
\end{equation}
Further, we denote by \(\mathbf{e}_a\) the frame dual to \(\boldsymbol{\theta}^a\), so that \(\mathbf{e}_a \intprod \boldsymbol{\theta}^b = \delta_a^b\), and by \(\star\) the Hodge star of \(\mathbf{g}\). Note that we use bold symbols to denote objects defined on \(M\).

Together with the projectors \(\vt: M \to \mathbb{R}\) and \(\vs: M \to \Sigma\) we define the time translation vector field \(\pdt\) such that it satisfies \(\pdt \intprod \ddd\vt = 1\) and \(\vs_*\pdt = 0\). It allows us to decompose the coframe, and any other differential form on \(M\), in the form
\begin{equation}
\boldsymbol{\theta}^a = \hat{\boldsymbol{\theta}}^a\,\ddd\vt + \vec{\boldsymbol{\theta}}^a\,, \quad
\hat{\boldsymbol{\theta}}^a = \pdt \intprod \boldsymbol{\theta}^a\,, \quad
\pdt \intprod \vec{\boldsymbol{\theta}}^a = 0\,.
\end{equation}
We denote the pullbacks of \(\hat{\boldsymbol{\theta}}^a\) and \(\vec{\boldsymbol{\theta}}^a\) to \(\Sigma\) by \(\hat{\theta}^a\) and \(\vec{\theta}^a\), i.e., we use non-bold font for objects defined on \(\Sigma\). The latter defines the metric and musical isomorphisms
\begin{equation}
q = \eta_{ab}\vec{\theta}^a \otimes \vec{\theta}^b\,, \quad
\tau^{\sharp} = q^{-1}(\cdot, \tau)\,, \quad
\zeta^{\flat} = q(\cdot, \zeta)
\end{equation}
for one-forms \(\tau\) and vector fields \(\zeta\). Writing \(\ast\) for the hodge star of \(q\), we define
\begin{equation}
\xi^a = -\frac{1}{6}\eta^{ae}\epsilon_{ebcd}\ast(\vec{\theta}^b \wedge \vec{\theta}^c \wedge \vec{\theta}^d)\,, \quad
\hat{\theta}^a = \alpha\xi^a + \beta \intprod \vec{\theta}^a\,,
\end{equation}
where the lapse function \(\alpha\) and shift vector field \(\beta\) are uniquely defined from the latter equation, which expands \(\hat{\theta}^a\) in the basis spanned by \(\xi^a\) and \(\vec{\theta}^a\). Finally, we use \(q\) and \(\xi^a\) to decompose any one-form \(\sigma_a\) into irreducible components
\begin{gather}
\trc{\sigma}_a = \frac{1}{3}q^{-1}(\vec{\theta}^b, \sigma_b)\vec{\theta}_a\,, \quad
\asy{\sigma}_a = \frac{1}{2}\left[q^{-1}(\vec{\theta}_a, \vec{\theta}^b)\sigma_b - q^{-1}(\vec{\theta}_a, \sigma_b)\vec{\theta}^b\right]\,,\nonumber\\
\sym{\sigma}_a = \frac{1}{2}\left[q^{-1}(\vec{\theta}_a, \vec{\theta}^b)\sigma_b + q^{-1}(\vec{\theta}_a, \sigma_b)\vec{\theta}^b\right] - \frac{1}{3}q^{-1}(\vec{\theta}^b, \sigma_b)\vec{\theta}_a\,, \quad
\vek{\sigma}_a = -\xi_a\xi^b\sigma_b\,,\label{eqn:irreddecomp}
\end{gather}
which satisfy \(\sigma_a = \trc{\sigma}_a + \asy{\sigma}_a + \sym{\sigma}_a + \vek{\sigma}_a\).

\section{Kinematic Hamiltonian of new general relativity}
The action of new general relativity\cite{Hayashi:1979qx} can be written in the form
\begin{equation}
S[\boldsymbol{\theta}^a, \boldsymbol{\omega}^a{}_b] = \int_M\boldsymbol{\mathcal{L}} = \int_M\left(C_T\TT^a \wedge \star\TT_a + C_V\VT^a \wedge \star\VT_a + C_A\AT^a \wedge \star\AT_a\right)\,.
\end{equation}
Here \(C_T, C_V, C_A\) are free constants, and the torsion components are given by
\begin{equation}
\VT^a = \frac{1}{3}\boldsymbol{\theta}^a \wedge (\mathbf{e}_b \intprod \mathbf{T}^b)\,, \quad
\AT^a = \frac{1}{3}\eta^{ab}\mathbf{e}_b \intprod (\eta_{cd}\boldsymbol{\theta}^c \wedge \mathbf{T}^d)\,, \quad
\TT^a = \mathbf{T}^a - \VT^a - \AT^a\,.
\end{equation}
From now on we will work in the Weitzenböck gauge \(\boldsymbol{\omega}^a{}_b \equiv 0\), without loss of generality.\cite{Blixt:2019mkt}. In order to derive the Hamiltonian, we first write the Lagrangian in the form \(\boldsymbol{\mathcal{L}} = \ddd\vt \wedge \hat{\boldsymbol{\mathcal{L}}}\), where the pullback \(\hat{\mathcal{L}}\) of the spatial part \(\hat{\boldsymbol{\mathcal{L}}}\) is given by
\begin{equation}
\begin{split}
\hat{\mathcal{L}} &= \frac{2C_T + C_V}{3}\left\{\alpha\dd\vec{\theta}^a \wedge \ast \dd\vec{\theta}_a - \frac{1}{\alpha}\left[\dot{\vec{\theta}}^a - \dd(\alpha\xi^a) - \Lie_{\beta}\vec{\theta}^a\right] \wedge \ast\left[\dot{\vec{\theta}}_a - \dd(\alpha\xi_a) - \Lie_{\beta}\vec{\theta}_a\right]\right\}\\
&+ \frac{C_T - C_V}{3}\left[\alpha(\dd\vec{\theta}^a \wedge \vec{\theta}_b) \wedge \ast(\dd\vec{\theta}^b \wedge \vec{\theta}_a) - \frac{1}{\alpha}\left(\dot{\vec{\theta}}^a \wedge \vec{\theta}_b + E^a{}_b\right) \wedge \ast\left(\dot{\vec{\theta}}^b \wedge \vec{\theta}_a + E^b{}_a\right)\right]\\
&+ \frac{C_A - C_T}{3}\left[\alpha(\dd\vec{\theta}^a \wedge \vec{\theta}_a) \wedge \ast(\dd\vec{\theta}^b \wedge \vec{\theta}_b) - \frac{1}{\alpha}\left(\dot{\vec{\theta}}^a \wedge \vec{\theta}_a + E^a{}_a\right) \wedge \ast\left(\dot{\vec{\theta}}^b \wedge \vec{\theta}_b + E^b{}_b\right)\right]\,.
\end{split}
\end{equation}
Here dots denote time derivatives, and we used the abbreviation
\begin{equation}
E^b{}_a = -\dd(\alpha\xi^b) \wedge \vec{\theta}_a + \alpha\xi_a\dd\vec{\theta}^b - (\Lie_{\beta}\vec{\theta}^b) \wedge \vec{\theta}_a\,.
\end{equation}
Note that there are no time derivatives of the temporal part \(\hat{\theta}^a\) of the tetrad, or equivalently the lapse \(\alpha\) and shift \(\beta\). The next step is to derive the canonical momenta \(p_a\) conjugate to the spatial tetrad components \(\vec{\theta}^a\). Varying the Lagrangian with respect to the velocities \(\dot{\vec{\theta}}^a\) and using the relation \(\delta_{\dot{\theta}}\hat{\mathcal{L}} = \delta\dot{\vec{\theta}}^a \wedge p_a\) we find
\begin{multline}
p_a = -\frac{2}{3\alpha}\Big\{(2C_T + C_V)\ast\left[\dot{\vec{\theta}}_a - \dd(\alpha\xi_a) - \Lie_{\beta}\vec{\theta}_a\right]\\
+ (C_T - C_V)\vec{\theta}_b \wedge \ast\left(\dot{\vec{\theta}}^b \wedge \vec{\theta}_a + E^b{}_a\right) + (C_A - C_T)\vec{\theta}_a \wedge \ast\left(\dot{\vec{\theta}}^b \wedge \vec{\theta}_b + E^b{}_b\right)\Big\}\,.
\end{multline}
To invert the relation between velocities \(v^a \equiv \dot{\vec{\theta}}^a\) and momenta, we split the momenta in the form \(\alpha\ast p_a = s_a - \pi_a\) into a part \(\pi_a\) linear in the velocities and \(s_a\) independent of the velocities. These are given by
\begin{equation}
\pi_a = \frac{2}{3}\left[(2C_T + C_V)v_a - (C_T - C_V)\vec{\theta}^{\sharp}_b \intprod \left(v^b \wedge \vec{\theta}_a\right) - (C_A - C_T)\vec{\theta}^{\sharp}_a \intprod \left(v^b \wedge \vec{\theta}_b\right)\right]
\end{equation}
and
\begin{equation}
s_a = \frac{2}{3}\left\{(2C_T + C_V)\left[\dd(\alpha\xi_a) + \Lie_{\beta}\vec{\theta}_a\right] + (C_T - C_V)\vec{\theta}^{\sharp}_b \intprod E^b{}_a + (C_A - C_T)\vec{\theta}^{\sharp}_a \intprod E^b{}_b\right\}\,.
\end{equation}
By applying the irreducible decomposition~\eqref{eqn:irreddecomp}, one finds the relations
\begin{equation}\label{eqn:velmom}
\vek{\pi}_a = \frac{2}{3}(2C_T + C_V)\vek{v}_a\,, \quad
\asy{\pi}_a = \frac{2}{3}(2C_A + C_T)\asy{v}_a\,, \quad
\sym{\pi}_a = 2C_T\sym{v}_a\,, \quad
\trc{\pi}_a = 2C_V\trc{v}_a\,.
\end{equation}
Note that depending on the vanishing or non-vanishing of the constant factors the terms \(\pi_a\) vanish, giving rise to another primary constraint, or do not vanish, and contribute to the momenta.\cite{Blixt:2018znp,Blixt:2019ene} The kinematic Hamiltonian \(\hat{\mathcal{H}}_0 = v^a \wedge p_a - \hat{\mathcal{L}}\) is then given by
\begin{multline}
\hat{\mathcal{H}}_0 = \frac{C_A - C_T}{3}\alpha\left[\xi_a\xi_b\dd\vec{\theta}^a \wedge \ast\dd\vec{\theta}^b - \dd\vec{\theta}^a \wedge \theta_a \wedge \ast(\dd\vec{\theta}^b \wedge \theta_b)\right] - C_T\alpha\dd\vec{\theta}^a \wedge \ast\dd\vec{\theta}_a\\
+ \frac{C_T - C_V}{3}\alpha(\vec{\theta}^{\sharp}_a \intprod \dd\vec{\theta}^a) \wedge \ast(\vec{\theta}^{\sharp}_b \intprod \dd\vec{\theta}^b) - (\alpha\xi^a + \beta \intprod \vec{\theta}^a)\dd p_a - \dd\vec{\theta}^a \wedge (\beta \intprod p_a)\\
+ \hat{\mathcal{H}}_0[\vek{p}] + \hat{\mathcal{H}}_0[\asy{p}] + \hat{\mathcal{H}}_0[\sym{p}] + \hat{\mathcal{H}}_0[\trc{p}] + \dd\left[(\alpha\xi^a + \beta \intprod \vec{\theta}^a)p_a\right]\,.
\end{multline}
Here the last term is a total derivative and hence does not contribute to the dynamics. The remaining terms in the last line depend on the presence or absence of constraints and read
\begin{gather}
\hat{\mathcal{H}}_0[\trc{p}] = \begin{cases}
0 & \text{for } C_V = 0\,,\\
-\frac{\alpha}{4C_V}\trc{c}_a \wedge \ast\trc{c}^a & \text{otherwise,}
\end{cases} \quad
\hat{\mathcal{H}}_0[\sym{p}] = \begin{cases}
0 & \text{for } C_T = 0\,,\\
-\frac{\alpha}{4C_T}\sym{c}_a \wedge \ast\sym{c}^a & \text{otherwise,}
\end{cases}\nonumber\\
\hat{\mathcal{H}}_0[\asy{p}] = \begin{cases}
0 & \text{for } 2C_A + C_T = 0\,,\\
-\frac{3\alpha}{4(2C_A + C_T)}\asy{c}_a \wedge \ast\asy{c}^a & \text{otherwise,}
\end{cases}\\
\hat{\mathcal{H}}_0[\vek{p}] = \begin{cases}
0 & \text{for } 2C_T + C_V = 0\,,\\
-\frac{3\alpha}{4(2C_T + C_V)}\vek{c}_a \wedge \ast\vek{c}^a & \text{otherwise,}
\end{cases}\nonumber
\end{gather}
where we used the abbreviations
\begin{align}
\vek{c}_a &= \ast\vek{p}_a - \frac{2}{3}(C_T - C_V)\xi_a\vec{\theta}^{\sharp}_b \intprod \dd\vec{\theta}^b\,, &
\sym{c}_a &= \ast\sym{p}_a\,,\nonumber\\
\asy{c}_a &= \ast\asy{p}_a - \frac{2}{3}(C_A - C_T)\vec{\theta}^{\sharp}_a \intprod \left(\vec{\theta}^b \wedge \dd\xi_b\right)\,, &
\trc{c}_a &= \ast\trc{p}_a\,.\label{eqn:ifconstr}
\end{align}
Note that the result is linear in lapse \(\alpha\) and shift \(\beta\), so that these quantities are Lagrange multipliers, corresponding to primary constraints arising from diffeomorphism invariance of the theory. If any of the constant factors in the relations~\eqref{eqn:velmom} vanishes, additional constraints appear, which force the corresponding term~\eqref{eqn:ifconstr} to vanish. This reproduces our result derived using tensor components.\cite{Blixt:2018znp,Blixt:2019ene}

\section{Conclusion}
We have derived the kinematic Hamiltonian and primary constraints of new general relativity in the language of differential forms. Our result agrees with the result of a previous calculation performed using tensor components.\cite{Blixt:2018znp,Blixt:2019ene} This consolidates our result, which is an important step towards counting the degrees of freedom in these theories. The latter will be achieved after deriving the constraint algebra, as it has been done for the teleparallel equivalent of general relativity.\cite{Okolow:2013lwa}

\section*{Acknowledgments}
The author thanks Daniel Blixt, Viktor Gakis and Christian Pfeifer for discussions. He gratefully acknowledges the full support by the Estonian Ministry for Education and Science through the Personal Research Funding project PRG356, as well as the European Regional Development Fund through the Center of Excellence TK133 ``The Dark Side of the Universe''.

\bibliographystyle{ws-ijmpa}
\bibliography{friedmann}
\end{document}